\documentclass[preprint,12pt]{elsarticle}

\usepackage[normalem]{ulem}  
\usepackage{color} 
\usepackage{graphicx}
\usepackage{amssymb}
\usepackage{amsmath}

\renewcommand\sout{\bgroup \color{red} \ULdepth=-.5ex \ULset}

\journal{Physics Letters B}

\begin{document}

\begin{frontmatter}


\title{$N\Omega$ dibaryon from lattice QCD near the physical point}

\author[a]{Takumi~Iritani}
\ead{takumi.iritani@riken.jp}
\author[b]{Sinya~Aoki}
\author[a,c]{Takumi~Doi}
\author[d]{Faisal~Etminan}
\author[a]{Shinya~Gongyo}
\author[c,a]{Tetsuo~Hatsuda}
\author[e]{Yoichi~Ikeda}
\author[f]{Takashi~Inoue}
\author[e]{Noriyoshi~Ishii}
\author[b]{Takaya~Miyamoto}
\author[b]{Kenji~Sasaki}
\author{(HAL QCD Collaboration)}
\address[a]{RIKEN Nishina Center (RNC), Saitama 351-0198, Japan}
\address[b]{Yukawa Institute of Theoretical Physics, Kyoto Univ., Kyoto 606-8502, Japan}
\address[c]{RIKEN Interdisciplinary Theoretical and Mathematical Sciences Program (iTHEMS), Saitama 351-0198, Japan}
\address[d]{Dept. of Phys., Faculty of Sciences, University of Birjand, Birjand 97175-615, Iran}
\address[e]{Research Center for Nuclear Physics (RCNP), Osaka Univ., Osaka 567-0047, Japan}
\address[f]{College of Bioresource Science, Nihon Univ., Kanagawa 252-0880, Japan}

\begin{abstract}
   The nucleon($N$)-Omega($\Omega$) system in the S-wave and spin-2 channel ($^5$S$_2$) is studied
  from the (2+1)-flavor lattice QCD with nearly physical quark masses
  ($m_\pi \simeq 146$~MeV and $m_K \simeq 525$~MeV).
  The time-dependent HAL QCD method is employed to convert the lattice QCD data of the
   two-baryon correlation function to the baryon-baryon potential and eventually to the
   scattering observables.   The $N\Omega$($^5$S$_2$) potential, obtained under the assumption that
     its couplings to the D-wave octet-baryon pairs are small, 
     is found to be  attractive in all distances and to produce  a quasi-bound state near unitarity:
   In this channel, the scattering length, the effective range and the binding energy  from QCD alone
   read  $a_0= 5.30(0.44)(^{+0.16}_{-0.01})$~fm,
   $r_{\rm eff} = 1.26(0.01)(^{+0.02}_{-0.01})$~fm,
   $B = 1.54(0.30)(^{+0.04}_{-0.10})$~MeV, respectively.
   Including the extra  Coulomb  attraction, the binding  energy of  $p\Omega^-$($^5$S$_2$)
   becomes $B_{p\Omega^-} = 2.46(0.34)(^{+0.04}_{-0.11})$~MeV.
  Such  a spin-2 $p\Omega^-$ state could be searched through
  two-particle correlations in $p$-$p$, $p$-nucleus and nucleus-nucleus collisions.
\end{abstract}

\end{frontmatter}

\newpage

\section{Introduction}
\label{sec:introduction}

Quest for dibaryons is a long-standing experimental and theoretical challenge in hadron physics \cite{Clement:2016vnl,Cho:2017dcy}.
 Among various theoretical attempts to study dibaryons, one of the recent  highlights
 is the (2+1)-flavor  lattice QCD simulations
near the physical point   ($m_\pi \simeq 146$~MeV and $m_K \simeq 525$~MeV) by HAL QCD Collaboration.  (For a recent summary,
 see Ref.\cite{Doi_2018}.)
 This enables us to make model-independent investigations of   the elusive  $H$-dibaryon,  originally proposed
  by the MIT bag model~\cite{Jaffe:1976yi},   on the basis of a coupled channel analysis of the lattice QCD data~\cite{Sasaki:2018mzh}.
  Also, the possible di-Omega ($\Omega\Omega$), originally proposed by the Skyrme model \cite{Kopeliovich:1990pp},
   has  recently been examined  in detail from the same lattice QCD data~\cite{Gongyo:2017fjb}.

Another interesting  candidate of the dibaryon is $N\Omega$ ($uudsss$ or $uddsss$)
in the $^5$S$_2$ channel.
 Since the Pauli exclusion does not operate among valence quarks and
 the color-magnetic interaction is attractive in the channel,
 it was predicted to  be a resonance below the $N\Omega$  threshold in the constituent quark model~\cite{Goldman:1987ma,Oka:1988yq}.
 Moreover, $N\Omega$($^5$S$_2$)  is expected to have relatively a small width since 
 its strong decay into octet baryons such as $\Lambda\Xi$ and $\Sigma\Xi$, which must have orbital D-wave,  would be 
kinematically suppressed.
 A pilot  (2+1)-flavor lattice QCD simulations
  with  a heavy pion mass ($m_\pi \simeq 875$~MeV)~\cite{Etminan:2014tya}
   suggests a short-range attraction
   between $N$ and $\Omega$ in the $^5$S$_2$ channel.
 Subsequently,  theoretical studies on the $N\Omega$ system
 \cite{Huang:2015yza,Morita:2016auo,Haidenbauer:2017sws,Sekihara:2018tsb,Garcilazo:2018gkb}
as well as  experimental  measurements  in
 relativistic heavy ion collisions~\cite{STAR:2018uho} have been reported.

The purpose of this Letter is to study  $N\Omega$($^5$S$_2$)
 on the basis of realistic (2+1)-flavor lattice QCD simulations near the physical point ($m_\pi \simeq 146$~MeV and $m_K \simeq 525$~MeV).
 As in the case of our previous pilot study~\cite{Etminan:2014tya},
  we  employ the HAL QCD method~\cite{Ishii:2006ec,Aoki:2009ji,HALQCD:2012aa} which allows us to
 extract the interaction between $N$ and $\Omega$
 from the spacetime dependence of the two-baryon correlation function on the lattice.

This paper is organized as follows.
 In Sec.~2, we introduce the HAL QCD method to extract the hadron interaction
from lattice QCD.  In Sec.~3, we summarize  the setup of our lattice QCD simulations near the physical point.
 In Sec.~4, we analyze the  $N\Omega$ system in $^5$S$_2$ channel in detail.
 Sec.~5 is devoted to  summary and concluding remarks.

\section{HAL QCD method}

Let us  consider the $N\Omega$($^5$S$_2$)
 characterized by the following
 two-baryon correlation function,
\begin{equation}
  C_{N\Omega}(\vec{r}, t)
  = \frac{1}{24}\sum_{\mathcal{R} \in \mathcal{O}} \sum_{\vec{x}}
  P_{\alpha\beta,\ell;\, \alpha'\beta',\ell'}^{\rm (s=2)} \langle 0 |N_\alpha(\mathcal{R}[\vec{r}] + \vec{x},t)
  \Omega_{\beta,\ell}(\vec{x},t) \overline{\mathcal{J}}^{N\Omega}_{\alpha'\beta',\ell'}(0) | 0 \rangle ,
  \label{eq:NBS_NOmega}
\end{equation}
with $\mathcal{J}^{N\Omega}$ being  the wall-type quark source.
The  interpolating operators for the nucleon and the $\Omega$-baryon are
\begin{eqnarray}
  N_\alpha(x) = \varepsilon_{abc}(u^{a\, T}(x)C\gamma_5 d^b(x))q_\alpha^c(x),
  \quad \Omega_{\beta,\ell}(x) = \varepsilon_{abc}s_\beta^a(x)(s^{b\, T}(x)C\gamma_\ell s^c(x)), \nonumber \\
& &    \label{eq:NOmega_op}
\end{eqnarray}
where  $\alpha$ and $\beta$ are Dirac indices,
$\ell$ is a spatial label of gamma matrices,
$a$, $b$, $c$ are the color indices and $C \equiv \gamma_4 \gamma_2$
and Dirac indices are restricted to the upper two components.
The summation over the cubic group element $\mathcal{R} \in \mathcal{O}$
leads to a projection onto the
S-wave state\footnote{Strictly speaking, this operation projects onto the $A_1^{+}$ state
which contains not only $L=0$ state but also  $L = 4, 6, \cdots$ states in the continuum theory.}.  
    On the other hand, the projection operator onto the spin-2 state 
    $P_{\alpha\beta,\ell;\, \alpha'\beta',\ell'}^{\rm (s=2)}$
    picks the diagonal elements of $S_z$ for the source and the sink
    and takes the average of $S_z = \pm 2, \pm 1, 0$ states,
which corresponds to $E^{+} \oplus T_2^{+}$
 irreducible representations of $SO(3,\mathbf{Z})$ \cite{Basak:2005ir,Dudek:2010wm}.

 In the present paper, we assume that the couplings of $N\Omega$($^5$S$_2$) to the D-wave octet-octet channels
  {\em below} the $N\Omega$ threshold ($\Lambda \Xi$ and  $\Sigma \Xi$) 
  are small.\footnote{A recent phenomenological study indicates that the volume integral of the $N\Omega$($^5$S$_2$) potential
    from the D-wave octet-octet channels below the $N\Omega$ threshold are insignificant $\sim$  10\%
  (Table IV of \cite{Sekihara:2018tsb}).}
If this holds true, the  $t$-dependence of 
      the correlation function $C_{N\Omega}(\vec{r}, t)$ would be dominated by $N\Omega$($^5$S$_2$)
    for certain range of $t$ before the octet-octet channels take over  at large $t$.   
We also assume that  the coupling to the  inelastic octet-decuplet channels (such as $\Lambda \Xi^\ast$ located just
   {\em above} the $N\Omega$ threshold) is sufficiently small in the range of $t$ adopted in the present paper.\footnote{The contributions from the $N\Omega$ ($^5$D$_2$), $N\Omega$ ($^3$D$_2$) 
  and   $\Lambda \Xi^\ast$($^5$S$_2$)  to the volume integral of the $N\Omega$($^5$S$_2$) potential
 are  found to be negligible $\sim \mathcal{O}(1)\%$  in a phenomenological study (Table IV of \cite{Sekihara:2018tsb}).}
 If such inelastic contributions are not negligible,  not only the $t$-dependence but also the non-locality of the single-channel  
 $N\Omega$($^5$S$_2$) potential  would become significant.
   To check the effects of the neglected states  mentioned above  in more detail,  
 the coupled-channel analysis of the HAL QCD method~\cite{Aoki:2012bb} is necessary. 
 We leave it as a future problem.

To extract the single-channel  $N\Omega$($^5$S$_2$) potential,
it is convenient to define the following ratio which we call the ``$R$-correlator'',
\begin{equation}
  R_{N\Omega}(\vec{r}, t) \equiv \frac{C_{N\Omega}(\vec{r},t)}{C_N(t) C_\Omega(t)},
  \label{eq:Rcorr_NOmega}
\end{equation}
 where $C_N(t)$ and $C_{\Omega}(t)$ are single-baryon correlators.
Below the inelastic threshold,  $ R_{N\Omega}(\vec{r}, t)$  can be shown to satisfy the integro-differential equation
with a non-local and energy-independent kernel $U(\vec{r}, \vec{r'})$  \cite{HALQCD:2012aa},
\begin{eqnarray}
 & & \left[- \frac{\partial}{\partial t}
 + \frac{1+3\delta^2}{8m} \frac{\partial^2}{\partial t^2} + \mathcal{O}(\delta^2\partial_t^3)  \right]
    R(\vec{r},t) = H_0 R(\vec{r},t) +  \int  U(\vec{r}, \vec{r'}) R(\vec{r'},t) d\vec{r'}, \nonumber \\
  \label{eq:t_dep_HAL}
\end{eqnarray}
with  $H_0\equiv -\nabla^2/2m $,  the reduced mass $m \equiv (m_N m_\Omega)/(m_N + m_\Omega)$ and
the asymmetry parameter $\delta \equiv (m_N - m_\Omega)/(m_N + m_\Omega)$.

The central potential in the leading-order (LO) analysis under the derivative expansion
$U(\vec{r}, \vec{r'}) = \sum_n V_n(\vec{r})\nabla^n\delta(\vec{r}-\vec{r'})$
is given by\footnote{Good convergence of this derivative expansion at low energies for the point-sink scheme
   has been demonstrated ~\cite{Iritani:2018zbt, Murano:2011nz}
 for the $NN$ and $\Xi\Xi$ channels  where the long-range part of the potentials are expected to be dominated by the 
 single-pion exchange.   Such a good convergence  in other channels without one-pion exchange such as 
  $N\Omega$  and $\Omega\Omega$ needs to be checked  explicitly in the future.}
\begin{equation}
  V_{\rm C}(r) =
  - \frac{H_0R_{N\Omega}(\vec{r}, t)}{R_{N\Omega}(\vec{r},t)}
  - \frac{(\partial/\partial t)R_{N\Omega}(\vec{r}, t)}{R_{N\Omega}(\vec{r},t)}
  + \frac{1+3\delta^2}{8m} \frac{(\partial^2/\partial t^2)R_{N\Omega}(\vec{r}, t)}{R_{N\Omega}(\vec{r},t)},
  \label{eq:V0_def}
\end{equation}
up to $\mathcal{O}(\delta^2 \partial_t^3)$-terms in the right hand side.
Spatial and temporal derivatives on the lattice at $(\vec{r},t)$ are calculated in central difference scheme
 using nearest neighbour  points.
If $R_{N\Omega}(\vec{r}, t)$ is dominated by a single state at large $t$, each term in the r.h.s.
of Eq.~(\ref{eq:V0_def}) should have  no $t$-dependence.  Such a single-state saturation, however, is not necessary to obtain $V_{\rm C}(r)$ in
  the time-dependent HAL QCD method as long as $R_{N\Omega}(\vec{r}, t)$ is dominated by
  the elastic states. In general, each term in the r.h.s. receives $t$-dependence which
   provides  ``signal'' instead of ``noise''   for $V_{\rm C}(r)$.
       (If there remains residual $t$-dependence  in $V_{\rm C}(r)$, it implies the necessity of the next-to-leading order of the derivative expansion 
        and/or the channel coupling to other states~\cite{Iritani:2018zbt, Murano:2011nz}.)  
        This is why  the data at moderate values of $t \sim$ 1~fm
are  sufficient to extract  the baryon-baryon interaction in HAL QCD  method.%
\footnote{This is in sharp contrast to  the so-called
     ``finite volume method'' for two-baryon systems. 
     It  requires strict ground state saturation, so that very large value of $t >$ 10~fm is necessary.
  For such  large $t$, however, no signal can be
     obtained  due to the explosion of statistical errors. See
     \cite{Iritani:2016jie,Iritani:2017rlk,Aoki:2017byw,Iritani:2018vfn} for explicit demonstration of this fact.
     Note also that this problem has been recognized in
     the studies of meson-meson scatterings~\cite{Briceno:2017max}
     and the use of the variational method~\cite{Luscher:1990ck}
     is known to be mandatory.
}   
   
\section{Lattice Setup}

Gauge configurations are generated by using the (2+1)-flavor lattice QCD
with the Iwasaki gauge action at $\beta = 1.82$
and the non-perturbatively $\mathcal{O}(a)$-improved
Wilson quark action with the six APE stout smearing with 
the smearing parameter $\rho = 0.1$ at nearly physical quark masses
($m_\pi \simeq 146$~MeV and $m_K \simeq 525$~MeV)~\cite{Ishikawa:2015rho}.
The lattice cutoff is $a^{-1} \simeq 2.333$~GeV ($a\simeq 0.0846$~fm)
and the lattice volume $L^4$ is $96^4$, corresponding to $La \simeq 8.1$~fm.
This  is sufficiently large volume  to accommodate  two baryons.
We employ the wall-type quark source
with the Coulomb gauge fixing. The periodic (Dirichlet)  boundary condition for the spatial
 (temporal) direction is imposed for quarks.
 The quark propagators are obtained by using the domain-decomposed
solver~\cite{Boku:2012zi,Terai:2013,Nakamura:2011my,Osaki:2010vj},
and the unified contraction algorithm is employed to calculate the correlation functions~\cite{Doi:2012xd}.

The forward and backward propagations are averaged
and the hypercubic symmetry on the lattice (4 rotations) are utilized
for each configuration.
414 configurations are available by picking up one per five trajectories:
For 207 configurations which are separated by ten trajectories,
48 source locations are used,
while 24 source locations are used for the rest (207 configurations),
and the  total number of  measurements read 119,232.
   The statistical errors are estimated by the jackknife method
   with 20 samples (bin size 5,952 measurements).
   We have checked that the bin size dependence is small
    by comparing the result with 40 samples (bin size 2,880 measurements).
The fit to the effective mass in the range
 $12 \le t/a \le 17$ for $N$ and $17 \le t/a \le 22$ for $\Omega$
 lead to $m_N= 954.7(2.7)$~MeV and $m_{\Omega}=1711.5(1.0)$~MeV.
These values are about 2\% heavier than physical values due to a slight
 difference of the present  quark masses from the physical point.

\section{Spin-2 $N\Omega$ potential}

\begin{figure}[t]
  \centering
  \includegraphics[width=0.6\textwidth,clip]{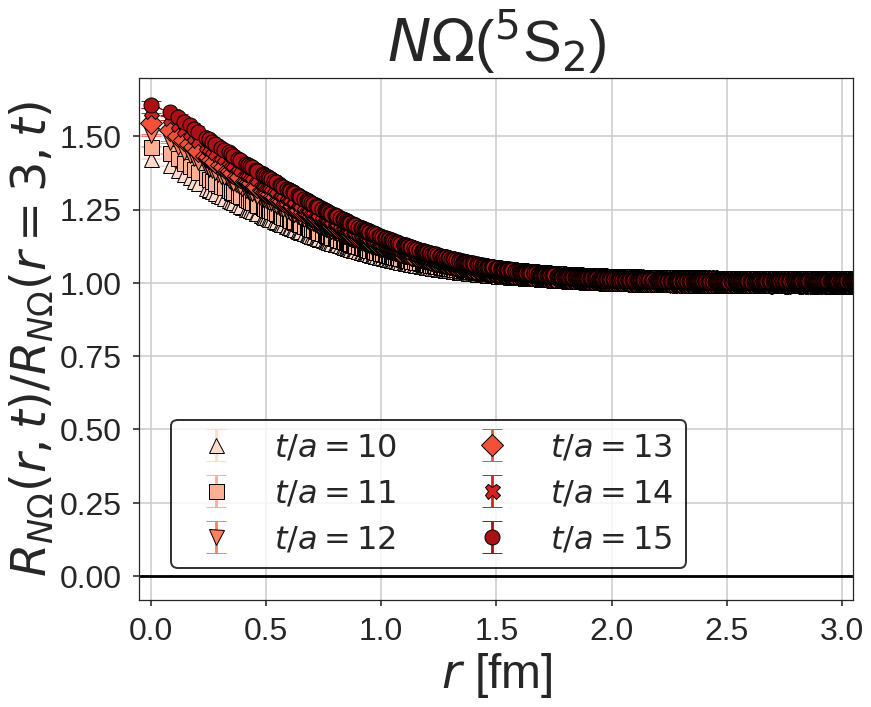}
  \caption{The rescaled $R$-correlator of the $N\Omega$($^5$S$_2$) in the range $t/a = 10-15$.}
  \label{fig:pot_breakup1}
\end{figure}

Shown in Fig.~\ref{fig:pot_breakup1} is the $R$-correlator defined by Eq.~(\ref{eq:Rcorr_NOmega})
in the range  $t/a = 10-15$, which are rescaled by the value of $r=3$~fm.
At large $r$, the $R$-correlator approaches a constant.
This implies that $V_{\rm C}(r)$ in Eq.~(\ref{eq:V0_def}) becomes a constant at long distance.
  At small $r$, the $R$-correlator increases with the second-order derivative in $r$ being always positive,  
 which implies that  there is an attractive
  potential at short distances.
 The weak $t$-dependence at small $r$
  indicates contributions from the elastic scattering states.
  As mentioned before, this $t$-dependence provides signal instead of noise.

To extract $V_\mathrm{C}(r)$ from the $R$-correlator, we choose $t/a = 11-14$  in order to reduce
the systematic uncertainties\footnote{Due to
  the presence of time derivatives up to $\mathcal{O}(\partial_t^2)$,  the actual  lattice data used in our analysis are in the interval
 $10 \leq t/a \leq 15$.}:    For smaller values of $t$,  the inelastic contribution starts to appear so that $V_\mathrm{C}(r)$
  remains non-vanishing even for large $r$.
  For larger values of $t$, it is difficult to control the systematic uncertainties of the fitting of the potential due to the large statistical errors.
 Note that  we take relatively larger values of  $t/a$
  to make accurate determination of  $m_{N}$ and $m_{\Omega}$, whose values agree with
  the effective masses at $t/a = 12$  in 1\%.

\begin{figure}[t]
  \centering
   \includegraphics[width=0.6\textwidth,clip]{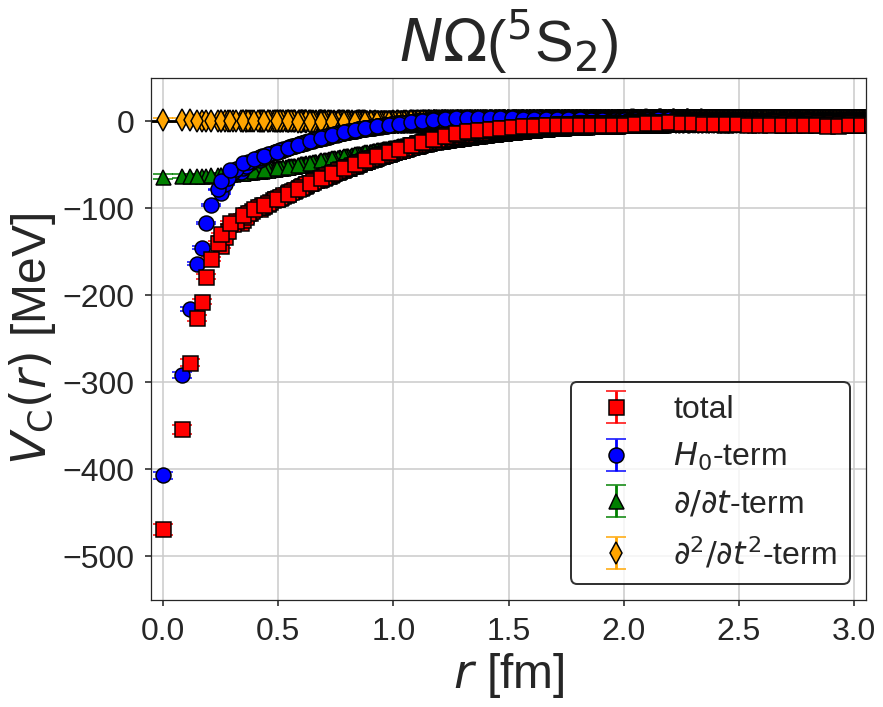}
  \caption{The central potential (red squares)  at $t/a = 12$ and its breakdown into
  the $H_0$-term (blue circles), the $\partial/\partial t$-term (green triangles)
  and the $\partial^2/\partial t^2$-term (orange diamonds).}
  \label{fig:pot_breakup2}
\end{figure}

  In Fig.~\ref{fig:pot_breakup2},
  $V_{\rm C}(r)$ as well as
  its breakdown into different components are shown for $t/a = 12$ as an example.
 First of all,  $V_{\rm C}(r)$ (red squares) is attractive everywhere. This is qualitatively consistent
 with the result in  our pilot study with heavy pion mass ($m_\pi \simeq 875$~MeV)~\cite{Etminan:2014tya}.
 Also, we found that the $H_0$-term (blue circles) is  dominant, yet the  $\partial/\partial t$-term (green triangles)
gives non-negligible $r$-dependent contribution. On the other hand, the $\partial^2/\partial t^2$-term (orange diamonds) is
  consistent with zero.

\begin{figure}
  \centering
   \includegraphics[width=0.47\textwidth,clip]{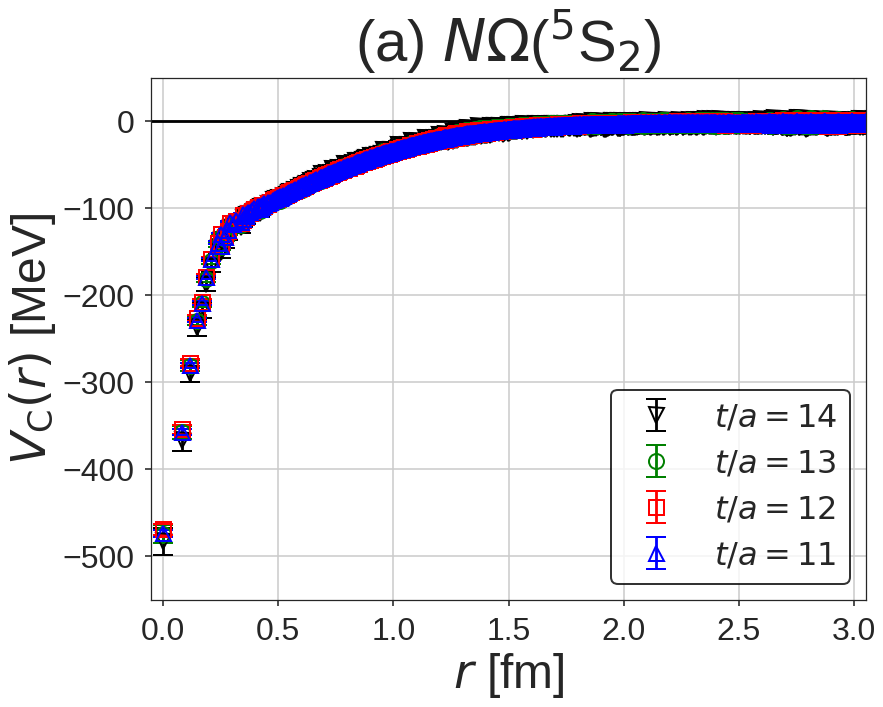}

  \includegraphics[width=0.47\textwidth,clip]{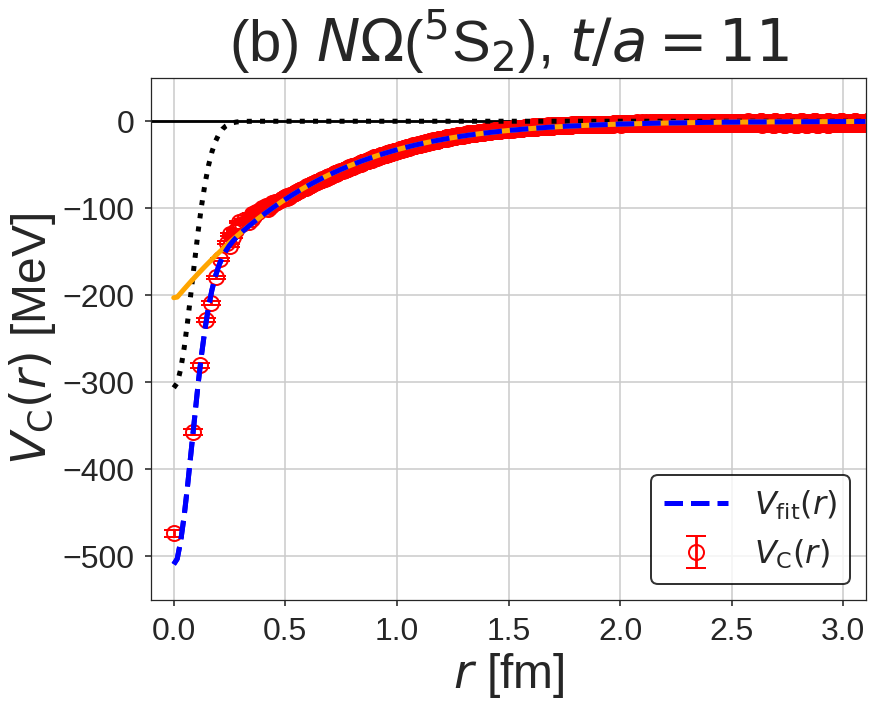}
  \includegraphics[width=0.47\textwidth,clip]{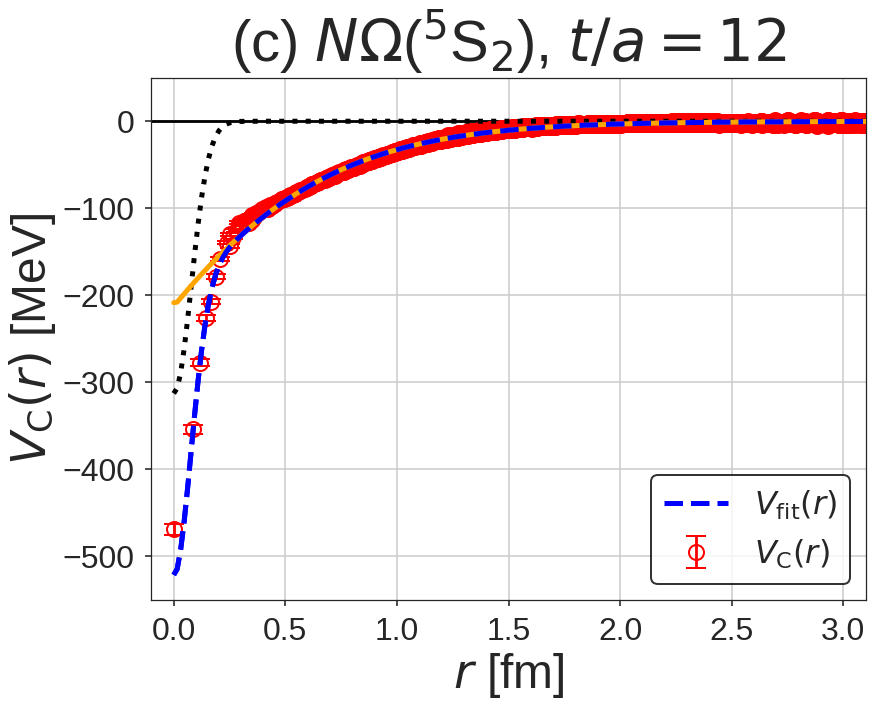}
  \includegraphics[width=0.47\textwidth,clip]{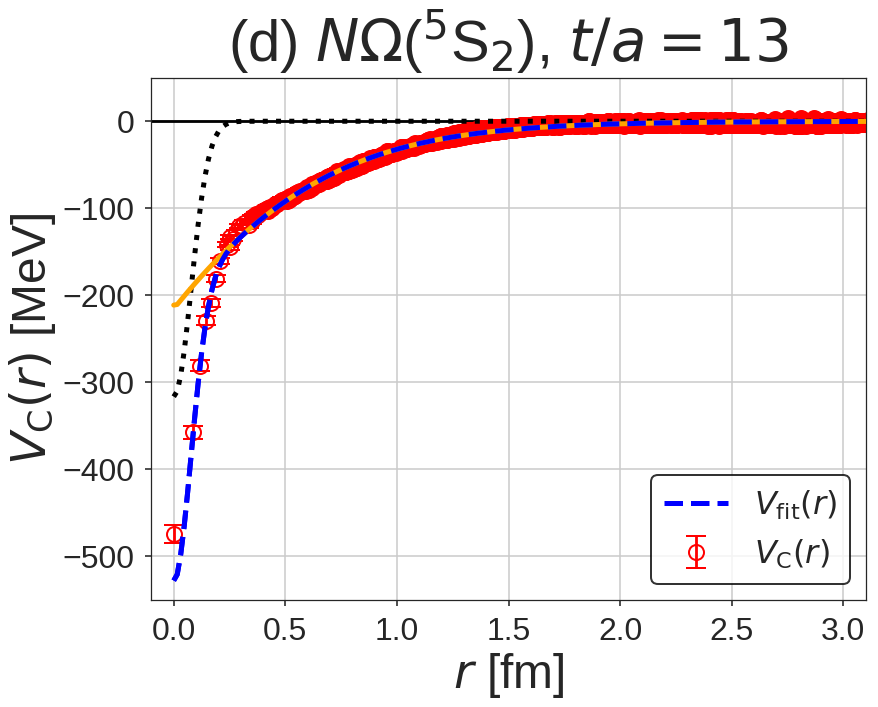}
  \includegraphics[width=0.47\textwidth,clip]{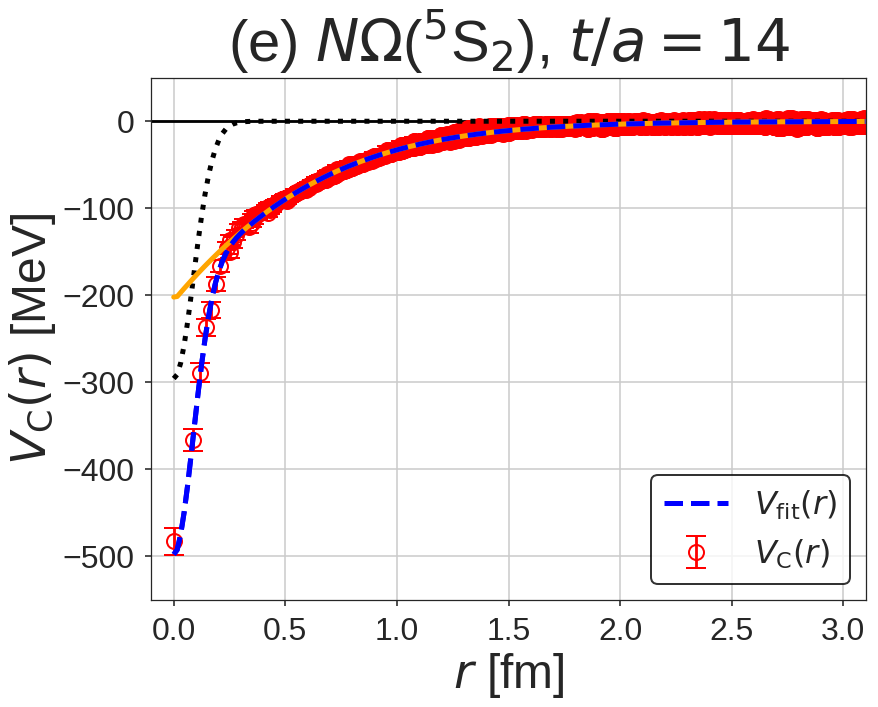}
  \caption{\label{fig:NOmegaPotFit}
  (a) The central potential $V_{\rm C}(r)$ of the $N\Omega(^5$S$_2)$ system at
  $t/a = 11$ (blue up-pointing triangles), 12 (red squares), 13 (green circles) and 14 (black down-pointing triangles).
  (b)
  The result of the fitting of $V_{\rm C}(r)$ (red circles) at $t/a=11$ by using $V_\mathrm{fit}(r)$ in Eq.~(\ref{eq:fit_func}).
  The black dotted (orange solid) line denotes the first (second) term in Eq.~(\ref{eq:fit_func}),
  and the blue dashed line is the sum of two terms. (c), (d) and (e) are the cases of $t/a=12$, 13 and 14, respectively.
}
\end{figure}

We summarize the central potential $V_{\rm C}(r)$ in Fig.~\ref{fig:NOmegaPotFit}(a) for $t/a = 11-14$.
These potentials are consistent with each other within statistical errors,
which is a necessary (but not sufficient) condition for the small coupling with the D-wave octet-octet states below the $N\Omega$ threshold
in the spin-2 channel.
 (Such a stability of the potential in the same range of $t$ in the spin-1 $N\Omega$ system is not found, which indicates 
  the strong coupling of the $N\Omega(^3$S$_1)$ state to the S-wave octet-octet states below threshold.)
 In the followings, we estimate the corresponding systematic errors
   as well as errors from the truncation of the derivative expansion and from the contamination of the inelastic states
   by utilizing the time dependence of the results.

To obtain observables such as the scattering phase shifts and binding energy, we fit the lattice  QCD potential
by  Gaussian +  (Yukawa)$^2$ with a form factor~\cite{Etminan:2014tya}:
\begin{equation}
V_\mathrm{fit}(r) = b_1 e^{-b_2 r^2} + b_3 \left( 1 - e^{-b_4 r^2} \right)^n
\left( \frac{e^{-m_\pi r}}{r} \right)^2.
\label{eq:fit_func}
\end{equation}
The (Yukawa)$^2$ form at long distance is motivated by the picture of
two-pion exchange between $N$ and  $\Omega$ with an OZI violating vertex~\cite{Sekihara:2018tsb}.
The pion mass in Eq.~(\ref{eq:fit_func}) is taken from  our simulation,
$m_{\pi}=146$~MeV, and we fit the data at $r < 3$ fm.   After trying  both $n=1$ and 2 in the form factor,
we  found that only $n=1$ can reproduce the short distance behavior of the
lattice potential, so that we will focus on the $n=1$ case below.
 The results of the fit  and the corresponding parameters are summarized in Fig.~\ref{fig:NOmegaPotFit}(b,c,d,e)
 and Table~\ref{tab:fit_params}, respectively\footnote{
       In order to examine the fit range dependence,
     we compare the fit with the data in $r < 2.5$~fm
     and that in  $r < 3$ fm
       by using the functional form of Eq.~(\ref{eq:fit_func}).
       The resultant scattering parameters are found to be consistent with each other within statistical errors.
     In addition, results for another functional form with three Gaussian 
       are found to be consistent with those obtained from Eq.~(\ref{eq:fit_func})
     within the statistical errors.
 }.

\begin{table}[h]
\centering
\begin{tabular}{l|rrrr}
  \hline \hline
  $t/a$ & $\ \ \ 11$ & $\ \ \ 12$ & $\ \ \ 13$ & $\ \ 14$ \\
  \hline
  $b_1$ [MeV] & $-306.5(5.5)$ & $-313.0(5.3)$ & $-316.7(9.4)$ & $-296(18)$ \\
  $b_2$ [fm$^{-2}$] & $73.9(4.4)$ & $81.7(5.4)$ & $81.9(8.4)$ & $64(16)$ \\
  $b_3$ [MeV$\cdot$fm$^2$]& $-266(32)$ & $-252(27)$ & $-237(43)$ & $-272(109)$ \\
  $b_4$ [fm$^{-2}$] &  $0.78(11)$ & $0.85(10)$ & $0.91(18)$ & $0.76(34)$ \\
    \hline \hline
  \end{tabular}
  \caption{The fitting parameters in Eq.~(\ref{eq:fit_func}) in physical unit with the statistical errors.
  }
  \label{tab:fit_params}
\end{table}

\begin{figure}
  \centering
  \includegraphics[width=0.47\textwidth,clip]{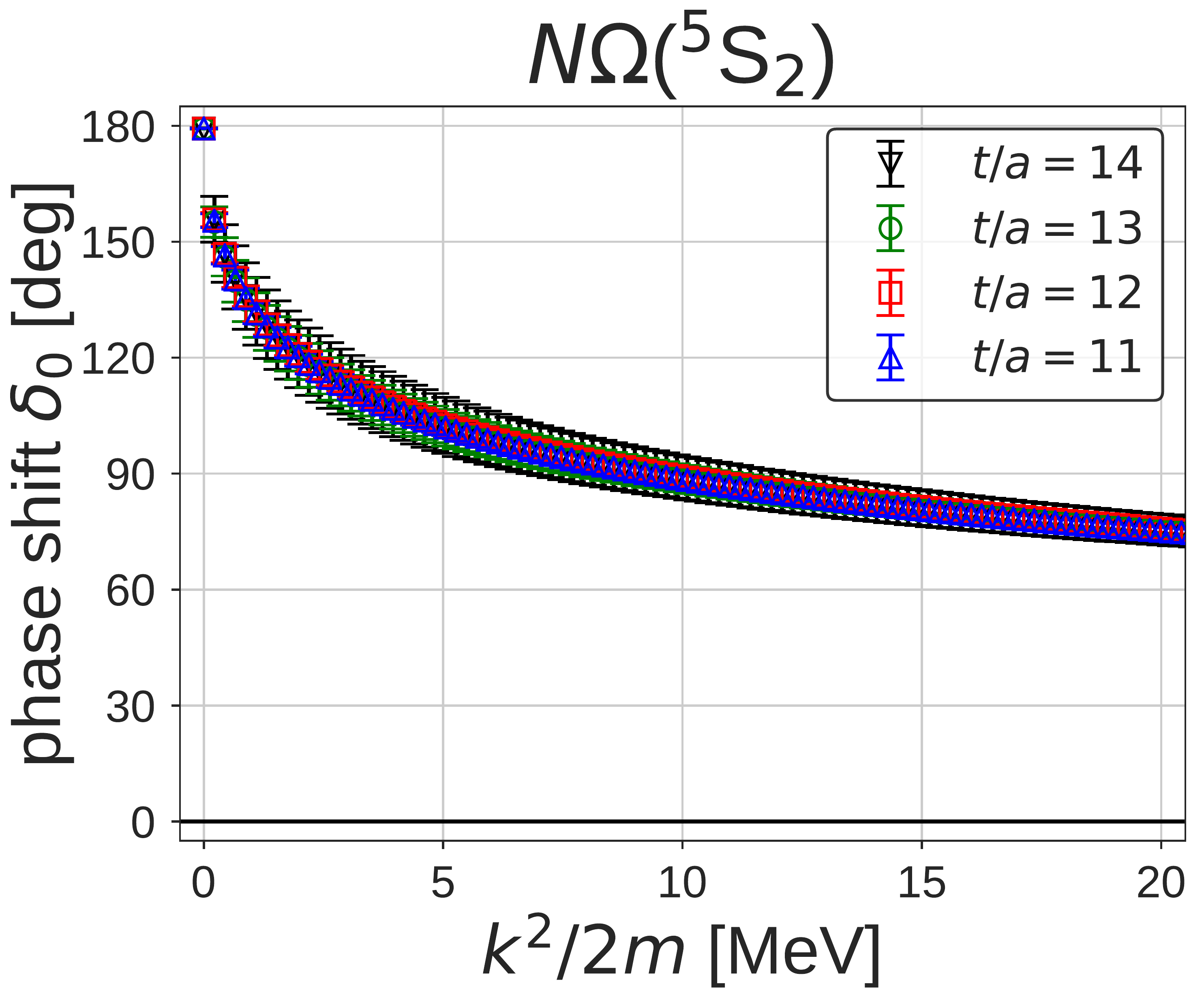}
  \includegraphics[width=0.47\textwidth,clip]{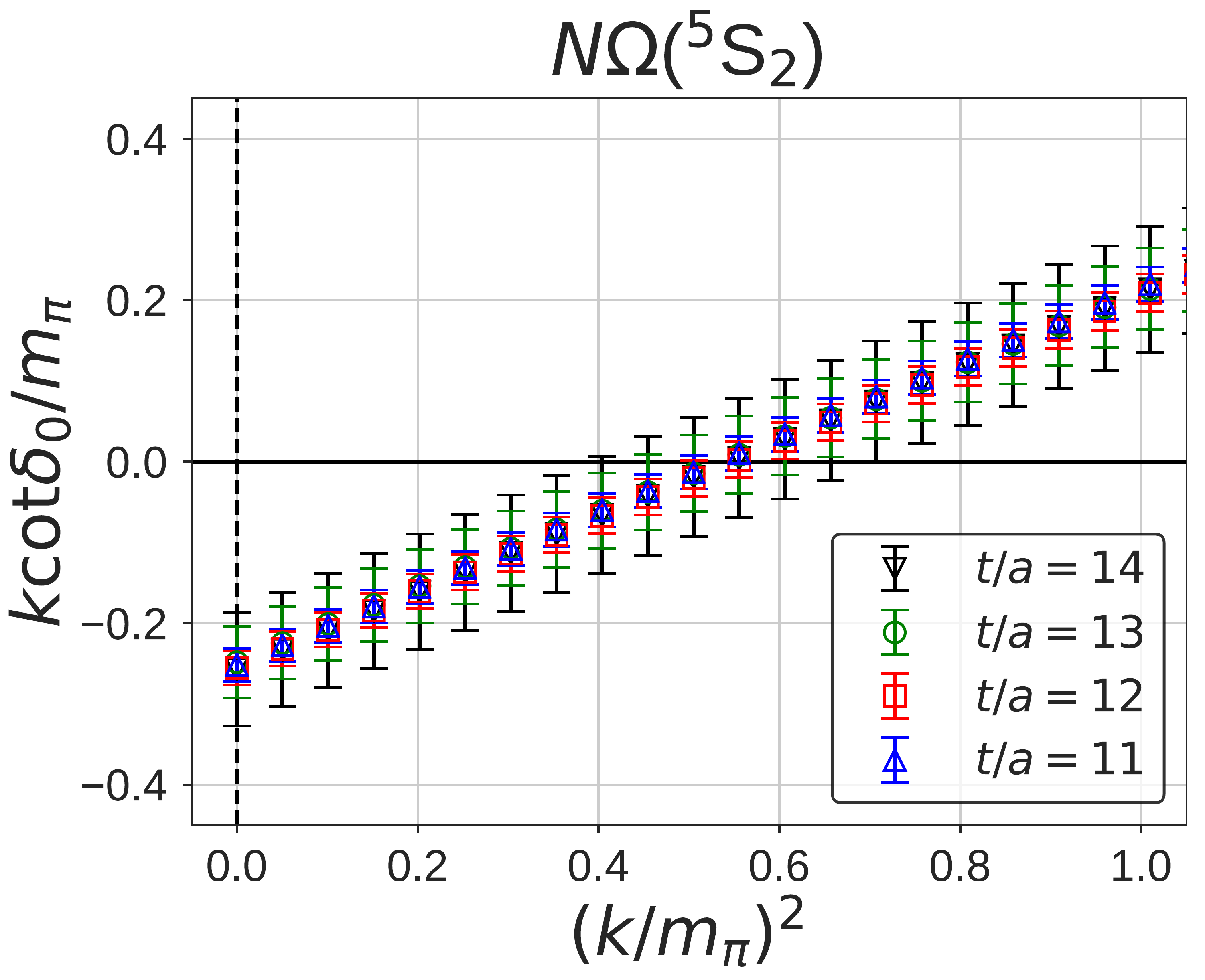}
  \caption{
    \label{fig:NOmegaPhaseShift}
    (Left) The S-wave scattering phase shifts $\delta_0$
    as a function of the kinetic energy, $k^2/2m$.
    (Right) $k\cot\delta_0/m_\pi$ as a function of $(k/m_\pi)^2$.
  }
\end{figure}

Shown in Fig.~\ref{fig:NOmegaPhaseShift}~(Left) is
the S-wave scattering phase shift $\delta_0$ as a function of the kinetic energy.
The values of $k\cot\delta_0$ are also shown in Fig.~\ref{fig:NOmegaPhaseShift}~(Right).
These results  for $t/a = 11$, 12, 13 and 14 are consistent  with each other within the statistical errors.
In the  $k \rightarrow 0$ limit, the phase shift approaches to $180^\circ$,
and the scattering length,\footnote{Here, the sign of the
scattering length is defined to be opposite to that in \cite{Etminan:2014tya}.}
$a_0 \equiv - \lim_{k\rightarrow 0}\tan\delta_0/k$,
becomes positive.
This implies the existence of a quasi-bound state of  $N\Omega$ in the $^5$S$_2$ channel.

The effective range expansion (ERE) of the phase shifts up to the next-leading-order (NLO)
 reads
\begin{equation}
  k\cot\delta_0 = - \frac{1}{a_0} + \frac{1}{2} r_\mathrm{eff}k^2 + O(k^4)
  \label{eq:ERE}
\end{equation}
with  $r_\mathrm{eff}$ being the effective range.  The ERE parameters $(a_0, r_\mathrm{eff})$
obtained from our phase shifts are found to be
\begin{equation}
  a_0 = 5.30(0.44)(^{+0.16}_{-0.01}) \ \mathrm{fm},
  \quad
  r_\mathrm{eff} = 1.26(0.01)(^{+0.02}_{-0.01}) \ \mathrm{fm},
  \label{eq:EREparam}
\end{equation}
where the central values and the statistical errors are estimated at $t/a = 12$, while
the systematic errors in the last parentheses are estimated from the central values for $t/a = 11$, 13 and 14.

\begin{figure}[h]
  \centering
  \includegraphics[width=0.7\textwidth,clip]{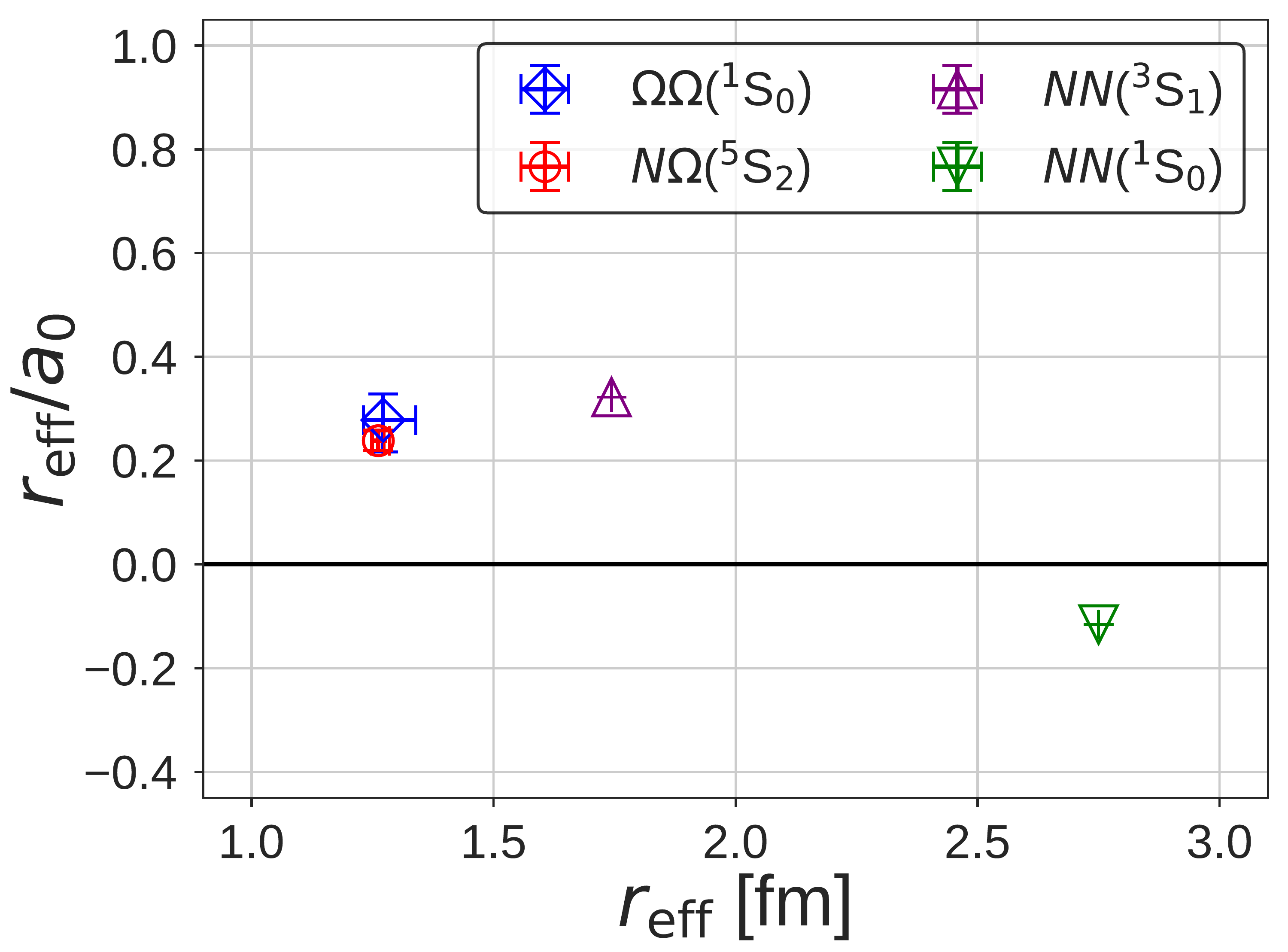}
  \caption{The ratio of the effective range $r_\mathrm{eff}$ and the scattering length $a_0$
  as a function of  $r_\mathrm{eff}$ for $N\Omega$($^5$S$_2$) (red circle)
  and $\Omega\Omega$($^1$S$_0$)~\cite{Gongyo:2017fjb} (blue diamond) on the lattice,
  as well as for $NN$($^3$S$_1$) (purple up-pointing triangle) and $NN$($^1$S$_0$) (green
  down-pointing triangle)~\cite{Hackenburg:2006qd} in experiments.}
  \label{fig:EREparam_dibaryons}
\end{figure}

 In Fig.~\ref{fig:EREparam_dibaryons}, the ratio $r_{\rm eff}/a_0$
  as a function of $r_{\rm eff}$ for $N\Omega$($^5$S$_2$)   is plotted together with the
 experimental values for $NN$($^3$S$_1$) (deuteron) and  $NN$($^1$S$_0$) (di-neutron)
  as well as lattice QCD value for $\Omega\Omega$($^1$S$_0$) (di-Omega)~\cite{Gongyo:2017fjb}.
 Small values of $|r_{\rm eff}/a_0|$ in all these cases indicate that these systems  are located
  close to the unitary limit.\footnote{The values in the fm unit are
 $ (a_0, r_\mathrm{eff})_{NN(^3\mathrm{S}_1)} = (5.4112(15), 1.7463(19))$,
  $ (a_0, r_\mathrm{eff})_{NN(^1\mathrm{S}_0)} = (-23.7148(43), 2.750(18))$ from the experiment~\cite{Hackenburg:2006qd},
  and $(a_0, r_\mathrm{eff})_{\Omega\Omega(^1\mathrm{S}_0)} = (4.6(6)(^{+1.2}_{-0.5}),
1.27(3)(^{+0.06}_{-0.03}))$ from the lattice QCD calculation~\cite{Gongyo:2017fjb}.}

The binding energy $B$ and the root mean square distance ($\sqrt{\langle r^2 \rangle}$)
of $N\Omega$($^5$S$_2$) are
 obtained by solving the Schr\"{o}dinger equation with the potential fitted to our lattice results:
\begin{equation}
  B = 1.54(0.30)(^{+0.04}_{-0.10}) \ \mathrm{MeV},
  \quad
  \sqrt{\langle r^2\rangle} = 3.77(0.31)(^{+0.11}_{-0.01}) \ \mathrm{fm}.
  \label{eq:BEparam}
\end{equation}
Although the $N$-$\Omega$  is attractive everywhere,  the binding energy is
as small as $\sim $1~MeV due to the short range nature of the potential.
 Accordingly,  the root mean square  distance  is comparable to the scattering length, indicating that
 the system  is  loosely bound like the   deuteron and the di-Omega.

In our pilot study ~\cite{Etminan:2014tya}, 
 we found $B = 18.9 (5.0) (^{+12.1}_{-1.8})$ MeV for heavy pion mass $m_{\pi} =  875$~MeV.
  The larger magnitude of $B$ than the present result  in  Eq.~(\ref{eq:BEparam})
  originates partly from the heavy  masses of $N$ and $\Omega$  in ~\cite{Etminan:2014tya}
   which   reduce the kinetic energy and thus increase the binding energy.  Another reason is that
   the short-range attraction for heavy pion  is relatively stronger.

\begin{figure}
  \centering
  \includegraphics[width=0.70\textwidth,clip]{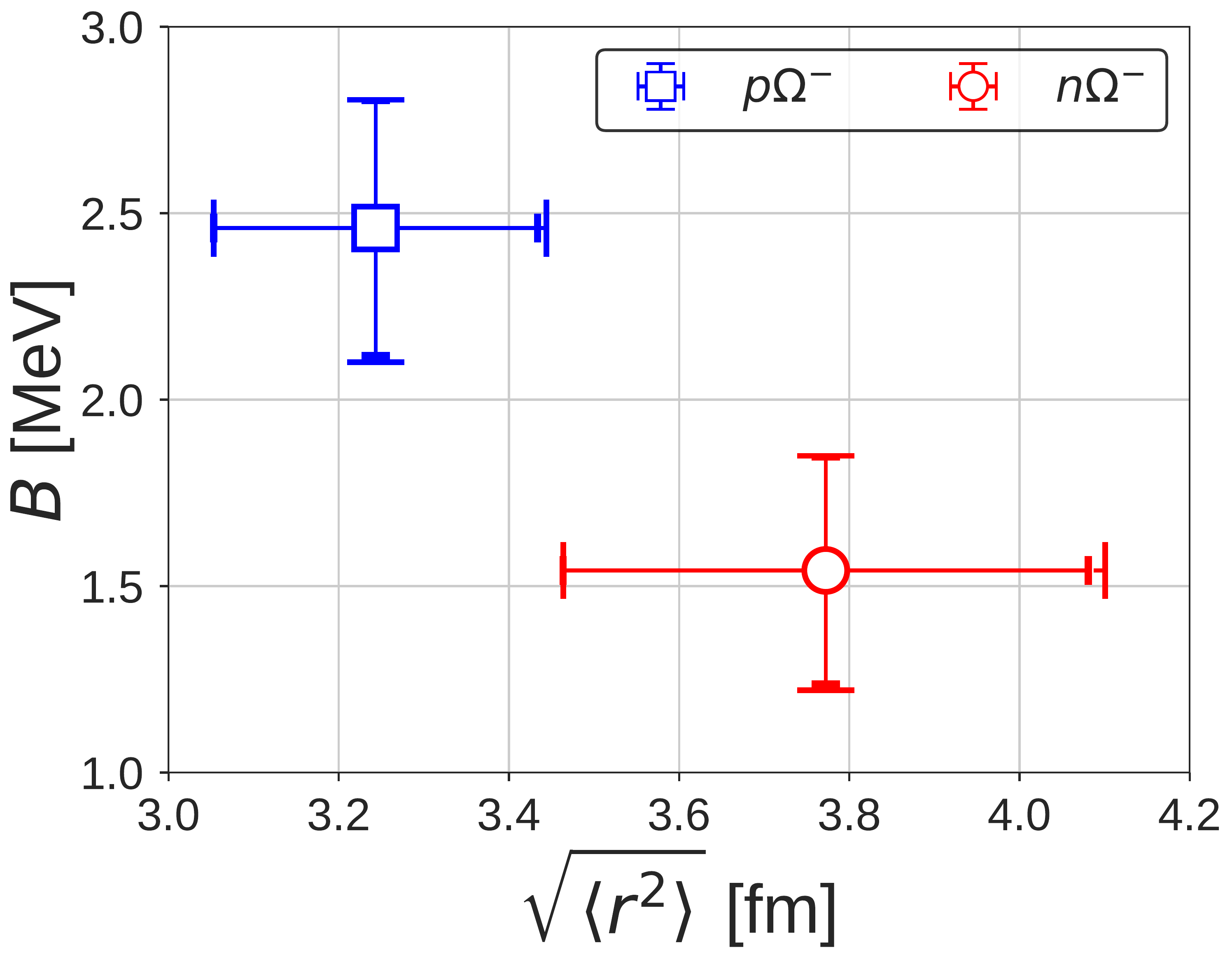}
  \caption{
    \label{fig:EREandBE}
    The binding energy $B$  and the root mean square  distance   $\sqrt{\langle r^2 \rangle}$
    for $n\Omega^-$ (red circle) and for $p\Omega^-$ (blue square).
    In both figures, inner bars correspond to the statistical errors, while
    the outer  bars are obtained by the quadrature of the statistical and systematic errors.
  }
\end{figure}

So far, we have not considered extra attraction in the $p\Omega^-$ system due to  Coulomb attraction.
By taking into account  the correction $V_{\rm C} (r) \rightarrow  V_{\rm C}(r) - \alpha/r$
with $\alpha \equiv e^2/(4\pi)=1/137.036$,  we obtain the observables,
\begin{equation}
  \label{eq:EREparam2}
  B_{p\Omega^-}  = 2.46(0.34)(^{+0.04}_{-0.11}) \ \mathrm{MeV}, \ 
  \sqrt{\langle r^2 \rangle}_{p\Omega^-}  = 3.24(0.19)(^{+0.06}_{-0.00}) \ \mathrm{fm}.
\end{equation}
These results  for  $p\Omega^-$($^5$S$_2$)
 are summarized in Fig.~\ref{fig:EREandBE}
  together with   $n\Omega^-$($^5$S$_2$) without Coulomb correction.

Before ending this section, let us briefly discuss other  possible systematic errors in
Eqs.~(\ref{eq:EREparam}), (\ref{eq:BEparam}) and (\ref{eq:EREparam2}).
The first one is the finite volume effect whose typical error
   would be  $\exp(-2m_{\pi}(L/2)) \simeq \exp(-6) \simeq 0.25 \%$ and is much smaller than the
   statistical errors in our simulation.  
   The second one is the finite cutoff effect,
   which is also expected to be small
   assuming the naive order estimate
   $(\Lambda a)^2 \leq 1$\% 
   with the non-perturbative $\mathcal{O}(a)$ improvement.
    The third systematic error is due to the slightly heavy hadron masses
     ($m_{\pi}=146$~MeV,  $m_{N}=955$~MeV and $m_{\Omega}=1712$~MeV).
By using the same parameter set for $t/a=12$ in  Table \ref{tab:fit_params}
 with $m_{\pi}=146$ MeV
 kept fixed but with physical baryon masses ($m_{p}=938$~MeV and $m_{\Omega^-}=1672$~MeV),
    we find less binding than   Eq.~(\ref{eq:EREparam2}) as expected:
      $B_{p\Omega^-} \simeq  2.18(32)$~MeV and $\sqrt{\langle r^2 \rangle}_{p\Omega^-}  \simeq  3.45(22)$~fm.
  On the other hand,
  if we additionally employ $m_\pi^\pm = 140$~MeV for the potential (see Eq.~(\ref{eq:fit_func})),
     we find more  bounding than Eq.~(\ref{eq:EREparam2}) due to smaller pion mass:
       $B_{p\Omega^-} \simeq  3.00(39)$~MeV and $\sqrt{\langle r^2 \rangle}_{p\Omega^-}  \simeq  3.01(16)$~fm.

\section{Summary}

In this paper, we have studied the $N$-$\Omega$ system in the $^5$S$_2$ channel, which is one of the promising candidates for
 quasi-stable dibaryon, from the (2+1)-flavor lattice QCD simulations with nearly  physical quark masses ($m_\pi \simeq 146$~MeV
 and $m_K \simeq 525$~MeV).
The $N$-$\Omega$ central potential in the $^5$S$_2$ channel obtained by the time-dependent HAL QCD method
  is found to be  attractive in all distances.
The scattering length and the effective range obtained by solving the Schr\"{o}dinger
equation using the resultant potential show  that  $N\Omega$($^5$S$_2$)  is close to unitarity similar to the
cases of the deuteron $(pn)$ and di-Omega ($\Omega\Omega$).
The binding energy of $p\Omega^-$ without (with) the Coulomb attraction is  about 1.5~MeV (2.5~MeV), which indicates
  the existence of  a shallow quasi-bound state below the $N\Omega$ threshold.
 In our simulation, we did not find a signature of the strong coupling between $N\Omega$($^5$S$_2$)
 and $\Lambda\Xi$ or $\Sigma\Xi$ in the D-wave state,
 while it remains to be an important future problem  to analyze  the coupled channel system with octet baryons, 
   $\Lambda\Xi$ and $\Sigma\Xi$.

 The $N\Omega$($^5$S$_2$) in the unitary regime can be studied in the two-particle correlation measurements
   in $p$-$p$ and $p$-nucleus and nucleus-nucleus collisions as suggested theoretically in \cite{Morita:2016auo} and
  experimentally reported  by  the STAR Collaboration at RHIC~\cite{STAR:2018uho}.
   Phenomenological analyses along this line on the basis of  the results in the present paper
 will be reported  elsewhere~\cite{Morita:2018}.

\section*{Acknowledgements}

We thank members of PACS Collaboration for the gauge configuration generation.
The lattice QCD calculations have been performed on the K computer at RIKEN
(hp120281, hp130023, hp140209, hp150223, hp150262, hp160211, hp170230),
HOKUSAI FX100 computer at RIKEN (G15023, G16030, G17002)
and HA-PACS at University of Tsukuba (14a-20, 15a-30).
We thank ILDG/JLDG~\cite{conf:ildg/jldg,Amagasa:2015zwb}
which serves as an essential infrastructure in this study.
We thank the authors of cuLGT code~\cite{Schrock:2012fj} for the gauge fixing.
This research was supported by
SPIRE (Strategic Program for Innovative REsearch),
MEXT as ``Priority Issue on Post-K computer''
(Elucidation of the Fundamental Laws and Evolution of the Universe) and JICFuS.
This work is supported by JSPS Grant-in-Aid for Scientific Research, No. 18H05236,
18H05407, 16H03978, 15K17667.
T.H. is grateful to the Aspen Center for Physics, supported in part by NSF Grants PHY1607611.
The authors thank T. Sekihara, K. Morita, and A. Ohnishi for fruitful discussions,
and H. Nemura and Y. Namekawa for useful comments.

\end{document}